\documentclass[twocolumn]{fairmeta}
\usepackage{amssymb}
\usepackage{cleveref} % cz: clever references
\usepackage{siunitx} % cz: scientific notation
\usepackage{bm} % cz: less bold bold
\usepackage{enumitem} % cz: customize padding for itemize
\usepackage{bbm} % dp: math-blackboard integers
\setlist[itemize]{topsep=3pt, partopsep=0pt, parsep=0pt, itemsep=3pt}
\usepackage[skip=2pt]{caption} % less whitespace around captions
\usepackage{subcaption} % new data dist figure
\usepackage{float}

\AtBeginDocument{%
  }

\renewcommand\normalsize{\fontsize{9}{11}\selectfont}
\begin{document}

\title{Enhancing Embedding Representation Stability in Recommendation Systems with Semantic ID}



\author[*, 1]{Carolina Zheng}
\author[\dagger]{Minhui Huang}
\author[\dagger]{Dmitrii Pedchenko}
\author[\dagger]{Kaushik Rangadurai}
\author[\dagger]{Siyu Wang}
\author[\dagger]{Gaby Nahum}
\author[\dagger]{Jie Lei}
\author[\dagger]{Yang Yang}
\author[\dagger]{Tao Liu}
\author[\dagger]{Zutian Luo}
\author[\dagger]{Xiaohan Wei}
\author[\dagger]{Dinesh Ramasamy}
\author[\dagger]{Jiyan Yang}
\author[\dagger]{Yiping Han}
\author[\dagger]{Lin Yang}
\author[\dagger]{Hangjun Xu}
\author[\dagger]{Rong Jin}
\author[\dagger]{Shuang Yang}

\affiliation[*]{Columbia University}
\affiliation[\dagger]{AI at Meta}

\contribution[1]{Work done during 2024 Internship at Meta}

\correspondence{
\email{
cz2539@columbia.edu, \{mhhuang, dmitripedchenko, krangadu, siyuw, gnahum12345, jielei, yzyang, tliu97, zutianluo, ubimeteor, dineshr, chocjy, yipinghan, ylin1, rongjinml, hangjunxu, shuangyang\}@meta.com
}
}

\metadata[Keywords]{Recommendation Systems, Content Understanding, Representation Learning, Vector Quantization}

% \renewcommand{\shortauthors}{Zheng et al.}

\abstract{

The exponential growth of online content has posed significant challenges to ID-based models in industrial recommendation systems, ranging from extremely high cardinality and dynamically growing ID space, to highly skewed engagement distributions, to prediction instability as a result of natural id life cycles (e.g, the birth of new IDs and retirement of old IDs). To address these issues, many systems rely on random hashing to handle the id space and control the corresponding model parameters (i.e embedding table). However, this approach introduces data pollution from multiple ids sharing the same embedding, leading to degraded model performance and embedding representation instability.

This paper examines these challenges and introduces Semantic ID prefix ngram, a novel token parameterization technique that significantly improves the performance of the original Semantic ID. Semantic ID prefix ngram creates semantically meaningful collisions by hierarchically clustering items based on their content embeddings, as opposed to random assignments. Through extensive experimentation, we demonstrate that Semantic ID prefix ngram not only addresses embedding instability but also significantly improves tail id modeling, reduces overfitting, and mitigates representation shifts. We further highlight the advantages of Semantic ID prefix ngram in attention-based models that contextualize user histories, showing substantial performance improvements. We also report our experience of integrating Semantic ID into Meta production Ads Ranking system, leading to notable performance gains and enhanced prediction stability in live deployments.
}

\newcommand{\cz}[1]{\textcolor{blue}{[cz: #1]}}
\newcommand{\mh}[1]{\textcolor{red}{[mh: #1]}}
\newcommand{\kr}[1]{\textcolor{orange}{[kr: #1]}}

\maketitle
% \pagestyle{plain}

% - where to put long retention experiment
% - more details on content understanding model (MARS)
% - can someone help with relevant citations in related work?
% - what new info are we incorporating? note the industry track requirement

% - split the segment analysis table
% - address blue comments
% - add more to the introduction
% - move discussion to where we are presenting results
% - move things from kaushik in the google doc
% - if i need help, use red color
% - overall description of MARS model

% - i still need to debug Table 1 (story won't change but there may be a bug)
% - also should figure out why pairwise distance is smaller for top 1000 clusters in Table 2; could expand this to a cluster composition section
% - also need to launch eval jobs and create figures

\section{Introduction}
Item recommendation can involve many signal-rich features, including categorical features corresponding to item IDs. The raw item IDs are usually mapped to embeddings, which are then further processed by deep learning-based model architectures such as the widely deployed Deep Learning Recommendation Model (DLRM) \citep{covington2016deep, naumov2019deep}. However, in industry-scale online settings, several key data-related challenges have emerged in learning item embedding representations. In particular, \textit{item cardinality}, the huge number of total items; \textit{impression skew}, the fact that only a few items comprise most user impressions or conversions \citep{milojevic2010power}; and \textit{ID drifting}, or the majority of items entering and leaving the system within short time periods \citep{gama2014survey}.

A popular and simple approach to learning embedding representations is random hashing, where raw item IDs are randomly hashed to share the same embedding \citep{zhang2020model}. Hashing is used due to the large item cardinality and system constraints on embedding table sizes. However, random hashing and ID drifting together lead to undesirable \textit{embedding representation instability} when the model is trained over long time periods. This is due to the nature of random hash collisions, which result in contradictory gradient updates to the embedding weights. Further, as the items in the system change over time from ID drifting, the learning from old items is lost and the embedding weights for new items are essentially random. This approach is ill-suited for items with few impressions, which are the majority of items due to impression skew.

To mitigate these drawbacks, a stable ID space is needed. A stable ID space ideally ensures that a learned embedding representation has a stable meaning as the model learns from more data. In this work, we investigate a recently proposed item representation approach called Semantic ID \citep{singh2023better, rajput2024recommender} as a candidate for a stable ID space. Semantic ID derives item IDs based on hierarchical clusters learned from the semantic similarity of items as given by their text, image, or video content. A given item's Semantic ID is then mapped to embedding representations via a parameterization scheme. Importantly, the ID space of Semantic ID is fixed \textit{a priori} and has semantic meaning -- meaning that it can address embedding representation instability. However, one challenge in using Semantic ID in recommendation modeling is defining a mapping from its cluster assignments to the embedding table.

The main contributions of this paper are:
\begin{itemize}
    \item Using experiments on a simplified version of Meta's production ads ranking model, we deepen the empirical understanding of how Semantic ID improves embedding representation stability. We further propose Semantic ID prefix-ngram, a novel token parameterization technique on top of Semantic ID that brings significant performance gains compared to the original Semantic ID introduced in \citep{singh2023better}.
    \item We characterize the item data distribution in terms of the number of items (item cardinality), the fact that most items have few impressions (impression skew), and the short item lifetime in the system (ID drifting) and explain their connection with embedding representation stability.
    \item We describe the productionization of Semantic ID prefix-ngram into both sparse and sequential features in Meta's production system. We show that adding these features brings online performance gains and improves online prediction stability.
\end{itemize}

In offline experiments on Meta's ads ranking data, we show that Semantic ID improves generalization and is less sensitive to distribution shift compared to random hashing. Confirming our hypothesis on impression skew, we find that most gains from Semantic ID come from the long tail of the item distribution. We show that by incorporating hierarchical cluster information, the proposed prefix-ngram is crucial to Semantic ID's effectiveness. We also demonstrate that semantic similarity translates to prediction similarity in both online and offline settings (\Cref{sec:representation_space} and \Cref{sec:online_semantic_similarity}). Further, Semantic ID results in outsized gains when incorporated in contextualizing models of users' item interaction history.

In an online setting, we describe the implementation of Semantic ID prefix-ngram features in Meta's production ads recommendation system, where they serve as the top sparse features by feature importance and result in $\mathbf{0.15\%}$ {\bf online performance gain}. Finally, we find that incorporating Semantic ID features significantly reduces the model's prediction variance for the same item. This is crucial to ensure advertisers' trust in Meta's recommendation system and to improve stability of the final item ranking.

The remaining sections are organized as follows: Section 2 explains related work. Section 3 provides an overview of the ranking model. Section 4 introduces Semantic ID and token parameterization. Section 5 explains the three item impression distribution challenges. Section 6 describes the offline experiments. Section 7 describes the productionization of Semantic ID at Meta and the online experiments. Section 8 concludes.

\section{Related Work}
\paragraph{Item representations in recommendation} Many modern deep learning recommendation models use trained embeddings to represent categorical (``sparse'') features \citep{covington2016deep, naumov2019deep, naumov2019dimensionality}. A simple solution to high item cardinality is to use random hashing \citep{weinberger2009feature}, but random hash collisions can be undesirable. One option is to modify the hashing procedure. Under this category, collision-free hashing \citep{liu2022monolithrealtimerecommendation} introduces individual embeddings for each item by dynamically free the memory of embeddings for retired items. Double hashing \citep{zhang2020model} utilizes two independent hash functions to reduce memory usage, but still has random collision. Learning to hash methods \citep{wang2017survey} focus on similarity preserving by training ML-based hash functions. There have also been works that address impression skew through contrastive learning or clustering \citep{yao2021self, chang2024cluster}; we view these as complementary approaches. We take a holistic approach of designing a stable ID space, to minimize the need for hashing and to address embedding representation shifting directly.

\paragraph{Stable embedding representation} Stable ID is inspired by tokenization approaches in NLP, which learn a fixed vocabulary of tokens to represent text in language modeling \citep{sennrich2015neural, kudo2018sentencepiece, devlin2018bert}. In designing a tokenization scheme for item recommendation, \citet{hou2023learning} proposes to vector-quantize the embeddings learned from an item content understanding model; \citet{qu2024tokenrec} introduce a masked vector-quantizer to transfer the learned representations from collaborative filtering models to a generative recommender. Semantic ID is introduced concurrently in \citep{singh2023better, rajput2024recommender}, which is based on \citep{hou2023learning} and uses an RQ-VAE for quantization, showing its benefits in generalization performance and sequential recommendation, respectively. In this work, we adapt Semantic ID as our stable ID method and analyze its effectiveness in addressing the three challenges in online item recommendation.
% \mh{TODO: broader related work}

% - sparse item mapping

% \paragraph{Semantic ID} \citet{hou2023learning} proposed the two-step paradigm of learning items' content embeddings and quantizing them into discrete codes, which are used to represent items in recommendation. \citet{rajput2024recommender} and \citet{singh2023better} utilize RQ-VAE as the vector quantizer, which we follow here. \citet{hou2023learning} and \citet{rajput2024recommender} apply this paradigm to sequential recommendation, where the task is to generate the next item based on the user history of item interactions. In our classification setting, the user history is only one source of features alongside sparse and dense user- and target item-side features. \citet{singh2023better} is the most closely related work to ours, as they show that Semantic ID improves generalizability over random hashing for target item and user history features in video recommendation. In this work, we go beyond reporting evaluation performance and analyze the reasons for and source of gains from Semantic ID.

% \paragraph{Generalization in recommendation} a) clustering, b) contrastive learning. \mh{I need help writing this part, as I'm not familiar with broader related work in recommendation.}

\section{Ranking Model Overview}
The recommendation problem is posed as a classification task, where a data point is the user- and item-side features associated with an item impression or conversion and a binary label indicating whether or not the user interacted or converted for that item. We now give a brief overview of the ranking model architecture.
%In \Cref{sec:system_overview_data}, we discuss the data, focusing on the item-based sparse features.

\subsection{Model}
\label{subsec:model}
The recommendation system follows a deep neural architecture based on the DLRM \citep{covington2016deep, naumov2019deep}. The model consists of three stacked sections. First is the information aggregation section, where the sparse (i.e., categorical), dense, and user history-based features are processed independently. The output of each of these modules is a list of embedding vectors. Second, these are concatenated into a single list which goes through the interaction layer, where dot products (or higher order interactions) are taken between all pairs of vectors. Third, the output of the interaction layer is transformed via an MLP to produce the logit score and a sigmoid is taken to output a probability. The model is trained using cross-entropy loss.

In the remainder of the paper we focus on the information aggregation section of the model. 

\paragraph{Embedding module} Let $I$ be the total number of raw IDs in the system and let  $[1..N]$ denote the integers from $1$ to $N$.

 The embedding table is a matrix $\mathbf{E} \in \mathbb{R}^{H \times d_m}$, where $d_m$ is the embedding dimension and $H$ is the number of embeddings. Let \mbox{$f=(f_1,\dots,f_G):[1..I] \to [1..H]^G$} be an embedding lookup function that maps a raw ID to $G$ embedding table row indices. Then for each raw ID $x \in [1..I]$, the sparse module looks up embedding rows $\mathbf{e}_{f_1}(x), \dots, \mathbf{e}_{f_G}(x)$ and produces a single output embedding via sum-pooling, $\mathbf{e}_f (x) := \sum_{i=1}^G \mathbf{e}_{f_i}(x)$. 

% \begin{equation}
%      \mathbf{e}_f 
%     (x) =
%     \sum_{i=1}^G \mathbf{E}^\intercal \mathbbm{1}_{f_i(x)},
%     \label{eq:sparse_module}
% \end{equation}
% where $\mathbbm{1}_{(\cdot)}$ is the one-hot encoding.
\paragraph{Sparse module}
A sparse feature is a set  $\mathbf{x} := \{x_1, \dots, x_n\}$ of  raw IDs. For instance, this could be a set of $n$ product category IDs a given item belongs to. We usually produce a single embedding $\mathbf{e}_f (\mathbf{x})$ by sum-pooling embeddings $\mathbf{e}_f (x_i)$ for constituent raw IDs. 

 %More generally, if a sparse feature is instead a list $\mathbf{x}=[x_1,...,x_k]$ of raw IDs, we represent it by sum-pooling individual embeddings $\mathbf{e}(\mathbf{x}) = \sum_{j=1}^k \mathbf{e}(x_j)$.

\paragraph{User history module} We model a user's item interaction history as a sequence of sparse features $\mathbf{x}^u := ( \mathbf{x}_1^u, \dots, \mathbf{x}_T^u )$ and the corresponding interaction timestamps. When working with these features, there are system constraints due to the number of items and the sequence length $T$. We include item interaction history for up to three months, which brings the item cardinality for the model to process to over one billion. It is important for the user history module to contextualize the sequence of features before they are further processed downstream. We describe the architecture below.

First, we use the sparse module 
%with a separately initialized embedding table
to embed each sparse feature $\mathbf{x}_i^u$ and obtain a learned timestamp embedding; the sum is $\mathbf{e}_f^u(\mathbf{x}_i^u)$. Let $\mathbf{X} = \left[ \mathbf{e}_f^ u(x_1^u); \dots; \mathbf{e}_f^u(x_T^u) \right]^\intercal \in \mathbb{R}^{T\times d_m}$ denote the resulting encoding.
We then contextualize this  sequence of embeddings via an aggregation module. We use one of the following three aggregation module architectures: Bypass, Transformer, and Pooled Multihead Attention (PMA), which are defined in \Cref{app:aggregation_modules}.

\subsection{Metrics}

\paragraph{Normalized Entropy} We measure model performance by normalized entropy (NE), defined as the model cross-entropy divided by the cross-entropy from predicting the data mean frequency of positive labels. The NE equation is
\begin{equation}
    \text{NE} = \frac{-\frac{1}{N} \sum_{i=1}^N (y_i \log(p_i) + (1-y_i)\log(1-p_i)}{-(p\log(p) + (1-p)\log(1-p))},
\end{equation}
where $N$ is the number of training examples, $y_i\in \{0,1\}$ is the label for example $i$, $p_i$ is the model prediction for example $i$, and $p=\frac{\sum_{i=1}^N y_i}{N}$. Lower is better.

%To directly compare a new model and a baseline model, we compute the NE percent gain, $(\text{NE}_{\text{new}} - \text{NE}_{\text{base}}) / \text{NE}_{\text{base}} \cdot 100$; a smaller number indicates better performance. Because we are ultimately interested in ranking items, we adjust away the effect of miscalibration and use calibration-free NE when reporting evaluation NE results. The calibration-free calculation is detailed in \Cref{app:calibration_free_ne}.

% what does it mean for practitioners
% our previous findings: EBFs bring outsized gains compared to target ad id.
%    - also in the Google video paper.
%    - can we analyze why this is? is it simply because EBFs have more features (200 vs. 1).

% H1: for tail ads, we could be underfitting rather than overfitting, since we only train for one epoch.
% H1/H2: cost for semantic ID: memorization (collision factor is higher)
% H1/H2: change granularity, aka collision rate: how does it affect memorization and generalization?
% H3: correlation btw metric that captures distribution shift and NE gain.

% attention maps: reduce self attention to itself.

% TODO: be more specific on experiments to run to prove each hypothesis. meet again w/ kaushik in a week.
%    - K: aim for industry paper. can probably be longer than 4 pages. hopefully no open source experiments are needed.

\section{Semantic ID and Parameterizations}
\label{sec:item_remapping_approaches}
The primary motivation for Semantic ID is to design an efficient clustering schema to represent items that allows knowledge sharing between items with shared semantics. Intuitively, if we have hundreds of ads about pizza that different users clicked on, we would want an example involving one of the ads to be informed by the other ads' representations. We craft the design of Semantic ID to potentially address the data-related challenges of \textit{item cardinality}, \textit{impression skew}, and \textit{ID drifting} described in \Cref{sec:system_overview_data}. %Semantics-based clusters may be more informative for recommendation than random clusters,
Compared to embedding representations based on random clusters, semantics-based representations will likely be more stable over time. Semantics-based clustering will also allow tail items to learn from more training examples. The learning from items that have left the system can also be utilized, and embedding weights for new items do not have to be learned from scratch. We investigate these hypotheses empirically in \Cref{sec:source_of_gains_exps}.

First, we give an overview of Semantic ID in \Cref{sec:semantic_id}. We then describe token parameterization in \Cref{sec:token_parameterization}. This step is crucial to incorporate Semantic ID into the recommendation model.

%ID drifting complicates the design of the mapping function $f(x)$, because we don't know \textit{a priori} which items will appear in future test data or their frequencies. A benefit of Semantic ID over the baselines we consider in \Cref{sec:baselines} is that if its collisions are meaningful for the prediction problem, the embeddings for new items that enter the system do not need to be learned from scratch.

\subsection{Overview}
\label{sec:semantic_id}
%The intuition behind Semantic ID is that because items share semantic information, we can define a mapping that creates semantic collisions. For example, if the task is to recommend products to a user who is an avid coffee fan, we may want to group coffee-related products together, and also have subcategories, e.g., for beans vs. filters vs. storage containers.
Semantic IDs are learned for items in two stages: first, apply a content understanding model to the items' text, image, or video to produce dense content embeddings. Then, train an RQ-VAE \citep{zeghidour2021soundstream} on the content embeddings to obtain a vector quantization for each item, which is represented as a sequence of coarse-to-fine discrete codes called the item's Semantic ID.
\begin{figure}[h]
    \centering
    \includegraphics[width=\linewidth]{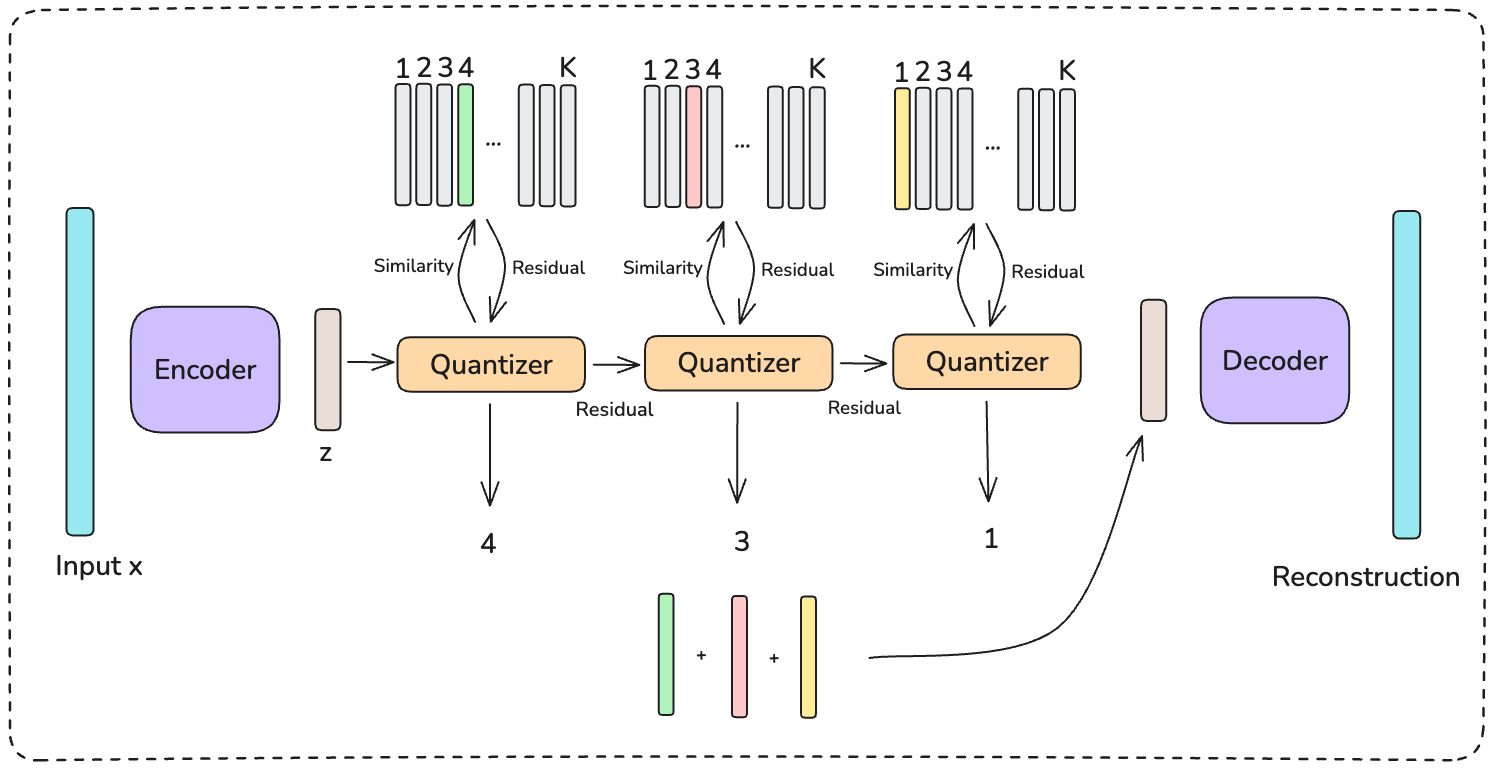}
    \caption{The RQVAE model with $L = 3$.}
    \label{fig:rqvae}
\end{figure}

Let $L$ be the number of layers (i.e., length of the sequence) and $K$ be the codebook size (i.e., number of clusters at each layer). RQ-VAE consists of an encoder that maps the content embedding $\mathbf{x} \in \mathbb{R}^D$ to a continuous latent representation, $\mathbf{z} \in \mathbb{R}^{D'}$, a residual quantizer that quantizes $\mathbf{z}$ into a series of discrete codes $\mathbf{c} := (c_1,\dots,c_L) \in K^L$, and a decoder that reconstructs $\mathbf{x}$ from $\mathbf{c}$. This is done by associating each layer $l$ with a codebook which is a set of $K$ vectors $\{\mathbf{v}^l_k\}_{k=1}^K$. The sequence of discrete codes is hierarchical: $c_l$ corresponds to the codebook vector $\mathbf{v}^l_{c_l}$ that approximates $\mathbf{r}_l$, the remaining residual from $\mathbf{z}$ after recursively applying the codebook vectors from layers $(l-1)$ to $1$, i.e.,
\begin{align}
    \mathbf{r}_l := \mathbf{z} - \sum_{i=1}^{l-1} \mathbf{v}^i_{c_i}, \quad
    c_l := \arg\min_c \| \mathbf{v}^l_c - \mathbf{r}_l \|_2.
\end{align}
In \Cref{sec:token_parameterization}, we provide more intuition on the nature of RQ-VAE's hierarchical clustering and how it informs the choice of token parameterization.

The RQ-VAE is trained using two loss terms, a reconstruction loss and a loss that encourages the residuals and codebook vectors to be close to each other,
\begin{align*}
    \mathcal{L}_{\text{RQ-VAE}}(\mathbf{x}) &= \|\mathbf{x} - \text{dec}(\mathbf{c})\|^2 \\
    &+ \sum_{l=1}^L \beta \| \mathbf{r}_l - \text{sg}(\mathbf{v}^l_{c_l}) \|^2 + \| \text{sg}(\mathbf{r}_l) - \mathbf{v}^l_{c_l} \|^2,
\end{align*}
where $\text{dec}(\mathbf{c})$ is the result of applying the decoder to the codes $\mathbf{c}$, $\text{sg}(\cdot)$ corresponds to the stop-gradient operator, and $\beta$ is a hyperparameter we set to $0.5$ in the experiments.
%For details on the RQ-VAE model, we refer the reader to $\dots$
A Semantic ID is defined as the sequence of discrete codes $(c_1,\dots,c_L)$ produced by the encoder and residual quantizer.
% Let $s(x): [1..I] \to K^L$ be the Semantic ID lookup that maps raw IDs to Semantic IDs. We must also specify a parameterization that maps a Semantic ID to embedding table rows, $p(\mathbf{c}): K^L \to [1..H]^G$. For the experiments, we use 3-layer prefix-gram, i.e.,
% \begin{equation}
%     p(\mathbf{c};H) = [c_1, (c_1+1)\cdot K+c_2, (c_1+1)\cdot K^2+(c_2+1)\cdot K+c_3\bmod H].
% \end{equation}

% Together, the Semantic ID map and the chosen parameterization define the mapping function $f_{\text{SemID}}(x; s, p) := p \circ s$.
% We present experimental results of alternative parameterization schemes in \Cref{app:sem_id_parameterizations}.

\subsection{Token Parameterization}

\label{sec:token_parameterization}

In our experiments, we use the same codebook size for each level, resulting in $K^L$ total clusters. An important feature of RQ-VAE is that it produces hierarchical clusters. Assuming $L=3$ for simplicity, a raw item ID is mapped to a sequence $(c_1, c_2, c_3)$.  The precision of vector quantization increases as one moves from the first token, $c_1$, to the deeper token $c_2$, and finally $c_3$. The first token $c_1$ represents the coarsest bucket: e.g., all ads related to food. The second token $c_2$ refines this information, e.g., $(c_1, c_2)$ may represent all ads related to pizza. The last token $c_3$ further refines this information, e.g., $(c_1,c_2,c_3)$ may represent all ads related to pizza and written in a specific language such as English.

Due to this, we can control the amount and the structure of the information that the recommendation model receives from Semantic ID. Notably, supplying the most fine-grained information (all possible $(c_1,c_2,\dots,c_L)$ tuples) is often not feasible due to high cardinality of the possible combinations. Hence, a tradeoff exists between the cardinality of the token parameterization and the amount of information the model receives from Semantic ID.

\begin{table}[h]
  \caption{Token parameterization techniques }
  \label{tab:token_parameterization_defs}
  \centering
  \footnotesize
  \begin{tabular}{c|ccc}
    Token Param & $p(c_1, \dots, c_L;H)$  \\ \midrule
    Trigram & $[K^2c_1+Kc_2+c_3]$  \\
    Fourgram & $[K^3c_1+K^2c_2+Kc_3+c_4]$  \\
    All bigrams & $[K^2 \times (i-1) + Kc_i+c_{i+1}, \quad \text{for i in [1 .. L-1]}]$ \\
    Prefix-ngram & $[\sum_{t = 1}^{i} K ^{i-t} (c_t+1)-1, \quad \text{for i in [1 .. n]} ]$ \\ \bottomrule
  \end{tabular}
\end{table}

Let $s(x): [1..I] \to K^L$ be the Semantic ID lookup that maps raw IDs to Semantic IDs learned by RQ-VAE. Keeping the hierarchical nature of the tokens in mind, we must specify a token parameterization that maps a Semantic ID to embedding table rows, $p(\mathbf{c};H): K^L \to [1..H]^G$. \Cref{tab:token_parameterization_defs} defines several possible parameterizations. When the Semantic ID cardinality is larger than the embedding size, a modulo hash function is applied. When there are multiple IDs, a shifting factor is added to avoid the collision between different positions. Among all the parameterization techniques, only Prefix-ngram consists of all possible tuples from different granularity.

\begin{table}[h]
  \caption{NE performance for different tokenization parameterizations}
  \label{tab:token_parameterization_ne}
  \centering
  \begin{tabular}{c|c|c}
    RQ-VAE $K \times L$  & Token Parameterization  & Train NE Gain \\ \midrule
    $[2048]\times 3$ &Trigram & $-0.028\%$ \\
    $[2048]\times 4$ &Fourgram & $-0.035\%$  \\
    $[2048]\times 4$ &All bigrams & $-0.091\%$ \\
    $[512]\times 3$ & Prefix-3gram  &  $-0.034\%$\\ 
    $[1024]\times 3$ & Prefix-3gram & $-0.097\%$ \\ 
    $[2048]\times 3$ &Prefix-3gram & $-0.141\%$ \\ 
    $[2048]\times 5$ &Prefix-5gram & $-0.208\%$ \\
    $[2048]\times 6$ &Prefix-6gram & $\bm{-0.215\%}$ \\\bottomrule 
  \end{tabular}
\end{table}

\Cref{tab:token_parameterization_ne} summarizes the model performance across different token parameterizations. We draw the following conclusions: i) Prefix-ngram is the best parameterization. This indicates that incorporating the hierarchical nature of the clustering in the embedding table mapping is necessary for effectiveness, as it allows for knowledge sharing among more items than a flat mapping; ii) Increasing the depth of Prefix-ngram improves NE performance; iii) increasing the RQ-VAE cardinality improves NE performance.

\section{Item Impression Distribution Issues}
\label{sec:system_overview_data}

In this section, we discuss the data distribution aspects that present challenges for recommendation modeling in Meta ads ranking and how we address them with the use of Semantic ID.

\paragraph{Item cardinality}
For certain features, such as the target item, the number of distinct items $I$ considered by the model can be much larger than a feasible embedding table size $H$ in the sparse module. In such cases the mapping function $f(\mathbf{x})$ introduces collisions: two or more raw IDs will map to the same row. The mapping function $f(\mathbf{x})$ is often chosen to be a simple hash. Since the initial raw IDs are randomly generated at the time of item creation, the resulting collisions will essentially be random. Such random collisions can negatively affect the resulting representation quality of embeddings and serve as an obstacle to effective knowledge sharing across items.
%Such random collisions could introduce difficulties in model training and generalization, as it is the main obstacle to knowledge sharing.
%We must design a mapping scheme that incorporates collisions, but is still effective for the prediction problem.

\begin{figure}[t]
\centering
\includegraphics[width=\linewidth]{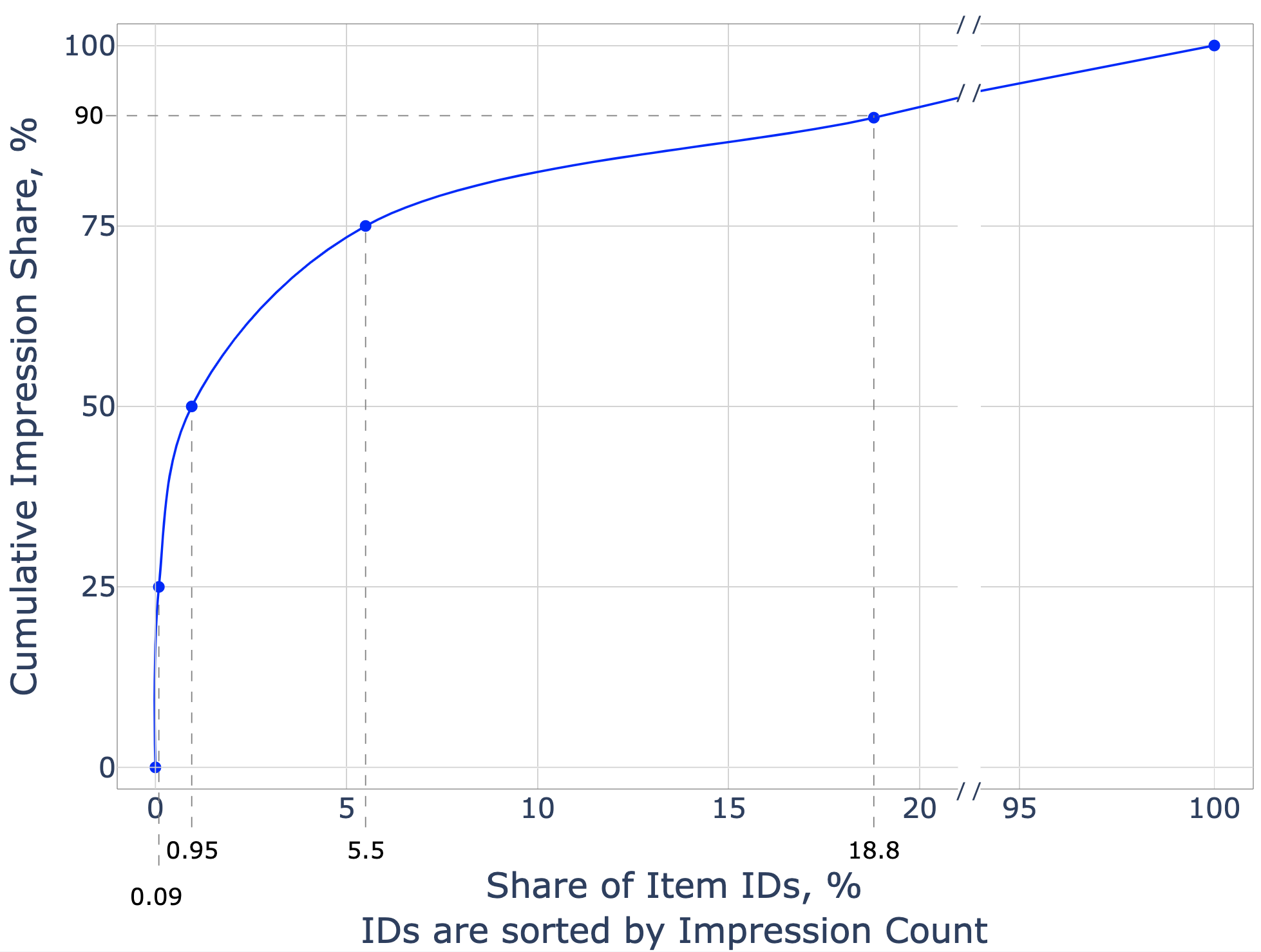}
 \caption{Impression Skew -- cumulative impressions as a function of the share of items considered. \textnormal{As items are sorted by the impression count, one sees that the majority of impressions comes from a fraction of most popular items.}}
\label{fig:impr_skew}
\end{figure}

\paragraph{Impression skew}

For the target item feature, the item distribution in the training data is highly skewed.  \Cref{fig:impr_skew} shows that in our system, a small percentage of items dominates the item impression distribution: when sorting the items by popularity, the top $0.1\%$ ``head'' items have $25\%$ of all item impressions,  the next $5.5\%$ ``torso'' items have $50\%$ of cumulative impressions, while the remaining $94.4\%$ ``tail'' items account for the remaining $25\%$ of impressions.

% \begin{figure}[h]
%     \centering
%     \includegraphics[width=0.6\linewidth]{figures/impression_skew.png}
%     \caption{Impression skew in the item distribution: most data comes from a few top items. The x-axis is items sorted by their number of training impressions, plotted against cumulative impressions.}
%     \label{fig:impression_skew}
% \end{figure}
%\cz{y-axis: cum. impressions, x-axis: items sorted by percentile}

As tail items have few training examples, it can be challenging to learn embedding representations $\mathbf{e}(\mathbf{x})$ that generalize well. Random hashing doesn't allow the head and torso items to effectively share knowledge with semantically similar tail items since the assignment of several items to a single embedding is random.

%effective knowledge sharing from head and torso items exacerbate this problem, as a single embedding must represent multiple items and tail items get fewer gradient updates. Therefore, it is difficult to encourage knowledge sharing among similar items in recommendation systems.

%One solution is to engineer collisions in the mapping function $f(x)$ to promote generalization; this is the intuition behind Semantic ID.

% \begin{figure}[h]
%     \centering
%     \includegraphics[width=0.75\linewidth]{figures/id_drifting.png}
%     \caption{ID drifting in the item distribution: items have a short lifetime in the system. The x-axis is the number of days elapsed. The left y-axis shows the percentage of items from day 0 still in the system, while the right y-axis shows the percentage of total items by day 30 that have entered the system so far.}
%     \label{fig:id_drifting}
% \end{figure}
%\cz{hazard curve: y-axis: percent survived, x-axis: number of days}

\paragraph{ID drifting}

\begin{figure}[t]
\centering
\includegraphics[width=\linewidth]{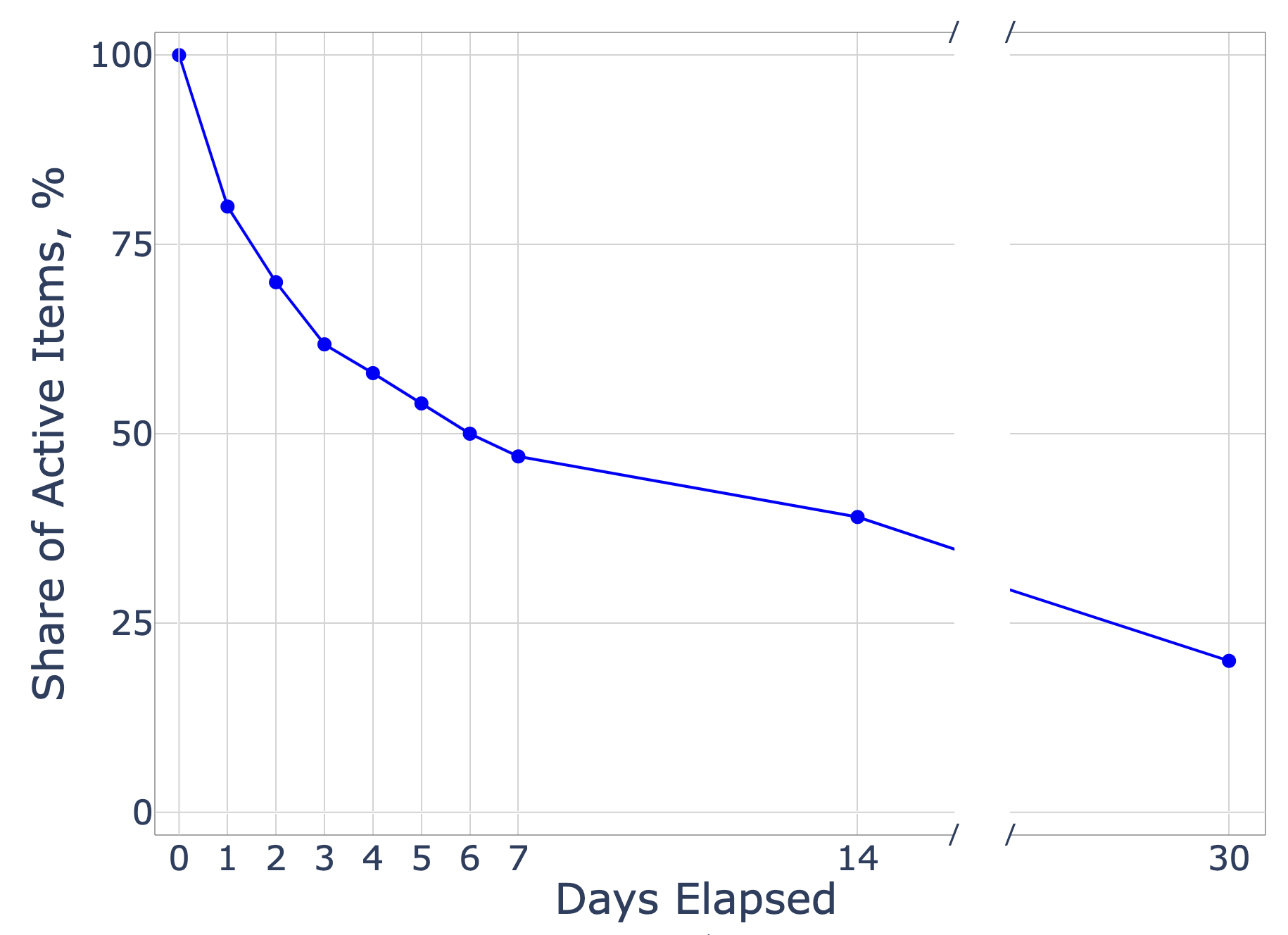}
\caption{ID Drift -- share of items that remain active in the initial corpus as a function of time. \textnormal{Half of the original corpus exits the system after 6 days. An equal number of new items enters the system, creating a severe item distribution drift.}}
\label{fig:id_drift}
\end{figure}

The existing item ID space is highly dynamic with large numbers of old items retiring (\Cref{fig:id_drift}) and new items entering the system. We call this item distribution shift in our system ``raw ID drifting.'' The raw ID drifting phenomenon stems from the nature of online recommendation systems, where new ads are created on a daily basis and most ads have a relatively short lifespan.

As a byproduct, a recommendation model based on random hashing experiences severe embedding representation drift over time: a given embedding $\mathbf{e}$ represents different items over time as items enter and exit the system.

\paragraph{Item representation with Semantic ID}
We hypothesize that switching from raw IDs to Semantic IDs can effectively address the issues above.

When an advertiser introduces a new ad $\mathbf{x}$ to the system and retires the previous one $\mathbf{y}$, the fine-grained content details of the new ad may be different from the retired one, but the broad semantic category of the product usually remains the same. Therefore, the new and retired ads' Semantic IDs will match (or at least share a prefix). Hence, the item impression distribution in Semantic ID space exhibits much less drift compared to the raw ID space as long as the broad semantic categories remain temporally stable.

Similarly, if a tail item $\mathbf{x}$ has similar content to a head or torso item $\mathbf{y}$, their Semantic IDs will match (or at least share a prefix). The resulting item impression distribution in Semantic ID space exhibits less skew compared to the raw ID space (see \Cref{app:semid_dist}).

In both of the cases above, the embeddings $\mathbf{e(x)}$ and $\mathbf{e(y)}$ will be equal (or similar if the Semantic IDs only share a prefix). This is a way for the model to transfer knowledge from item $\mathbf{y}$ with many training examples to item $\mathbf{x}$. Summarizing, temporal stability of semantic concepts results in stability of Semantic ID encoding, which mitigates embedding representation instability for the model.

\begin{table*}[t]

  % \centering
  \caption{Performance of three item representation approaches over various item segments.}
  \vspace{0.1cm}
  \begin{subtable}{0.61\textwidth}
  \caption{Evaluation NE (lower is better). \textnormal{Semantic ID enables knowledge transfer to tail and new cold start items.}}
    \label{tab:segment_analysis_generalization}
    \footnotesize
  \begin{tabular}{cc|ccc|cc}
    & & \multicolumn{3}{c|}{Eval NE} & \multicolumn{2}{c}{SemID NE Gain vs.} \\ 
    Cum. Exs. & Item Percentile & RH & IE & SemID & RH & IE \\ \midrule
    25\% (Head) & $0.1$ & 0.80105 & $\bm{0.80101}$ & 0.80108 & $0.00\%$ & \:$0.01\%$ \\
    75\% (Torso) & $5.6$ & 0.83589 & 0.83583 & $\bm{0.83580}$ & $-0.01\%$ & $-0.00\%$ \\
    100\% (Tail) & $100$ & 0.83904 & 0.83886 & $\bm{0.83872}$ & $-0.04\%$ & $-0.02\%$ \\ \midrule
    \multicolumn{2}{c|}{Items Seen in Training} & 0.82626 & 0.82612 & $\bm{0.82600}$ & $-0.03\%$ & $-0.02\%$ \\ 
    \multicolumn{2}{c|}{New Items} & 0.83524 & 0.83453 & $\bm{0.83180}$ & $-0.41\%$ & $-0.33\%$ \\ 
    \multicolumn{2}{c|}{All Items} & 0.82663 & 0.82645 & $\bm{0.82621}$ & $-0.05\%$ & $-0.03\%$ \\ \bottomrule
  \end{tabular}
  \end{subtable}
  \hspace{1cm}
  \begin{subtable}{0.3\textwidth}
  \caption{Sensitivity to distribution shift: 
    $\textbf{NE}\mathbf{[t_0, t_1]} - \textbf{NE}
    \mathbf{[t_0 + 42h, t_1 + 42h]}.$ Lower is better.}
  \label{tab:segment_analysis_drifting_gap}
  \vspace{0.1cm}
  \footnotesize
 \begin{tabular}{c|ccc}
    %& \multicolumn{3}{c}{NE(Eval) - NE(Train)} \\ 
    Cum. Exs. & RH & IE & SemID \\ \midrule
    25\% (Head) & 0.0057 & 0.0065 & 0.0059 \\
    75\% (Torso) & 0.0087 & 0.0075 & 0.0076 \\
    100\% (Tail) & 0.0128 & 0.0103 & 0.0106 \\ \midrule
    All Items & 0.0083 & 0.0074 & $\bm{0.0073}$ \\ \bottomrule
  \end{tabular}
  \end{subtable}
\end{table*}

\section{Offline Experiments}
% \begin{table*}[h]
%   \caption{Semantic ID NE \% gain by segment. \cz{todo: separate out eval on train from the eval, change to use NE loss}}
%   \label{tab:segment_analysis}
%   \centering
%   \begin{tabular}{cc|ccc|ccc}
%     & & \multicolumn{3}{c|}{vs. RH} & \multicolumn{3}{c}{vs. IE} \\ 
%     Cum. Exs. & Item Percentile & Train (mid 6h) & Train (last 6h) & Eval. & Train (mid 6h) & Train (last 6h) & Eval. \\ \midrule
%     25\%   & 0.09\% & 0.03\% & 0.03\% & 0.00\% & 0.05\% & 0.03\% & 0.00\% \\
%     50\%   & 0.86\% & -0.08\% & 0.12\% & -0.02\% & 0.14\% & 0.15\% & -0.01\% \\
%     75\%   & 4.68\% & -0.01\% & 0.07\% & -0.01\% & 0.07\% & 0.07\% & 0.00\%  \\
%     90\%   & 13.3\% & -0.14\% & 0.18\% & -0.05\% & 0.30\% & 0.32\% & -0.02\%  \\
%     100\%   & 100\% & -0.12\% & 0.45\% & -0.12\% & 0.78\% & 1.15\% & -0.06\% \\ \midrule
%     \multicolumn{2}{c|}{existing items} & -0.06\% & 0.16\% & -0.04\% & 0.21\% & 0.30\% & -0.02\% \\ \multicolumn{2}{c|}{new (cold start)} & --- & --- & -0.42\% & --- & --- & -0.25\% \\ \multicolumn{2}{c|}{all} & -0.06\% & 0.16\% & -0.06\% & 0.21\% & 0.30\% & -0.03\% \\ \bottomrule
%   \end{tabular}
% \end{table*}
\label{sec:source_of_gains_exps}
To investigate our hypotheses about the advantages of Semantic ID over baseline item representation approaches, we conduct a suite of offline experiments.
%We analyze the results in the context of our research questions, how Semantic ID addresses the challenges in item recommendation and user history modeling.

We use a simplified version of Meta's production ads ranking model, keeping all the dense features and the user's item interaction history, but including only the target item in the sparse module (and removing $O(100)$ other sparse features). We train on production user interaction data from a four-day time period, processing the training data sequentially and training for a single epoch. 
%This takes about 6 hours to train on 32 NVIDIA A100 80GB GPUs. 
We evaluate the model on the first six hours of the next day of data.
%, 8/11-8/15/24 and 8/25-8/29/24, with $\approx$30B examples each. There are $\approx$50M distinct target items and $\approx$1B distinct items in the user history sequences.

\subsection{Baselines}
\label{sec:baselines}
In \Cref{sec:system_overview_data}, we outlined data-related challenges and opportunities in designing a good embedding lookup function $f(\mathbf{x})$ for embedding-based item representations. We describe two baseline approaches, individual embeddings (IE) and random hashing (RH), and compare those to Semantic ID (SemID).

\paragraph{Individual embeddings} Each raw ID gets its own embedding table row, $I=H$ and $f_{\text{IE}}(x) :=x$. While unrealistic in production scenarios due to system constraints, we consider this model for illustration purposes. During evaluation, IDs that are not seen during training are mapped to a randomly initialized untrained embedding.

\paragraph{Random hashing} In the case where $I \approx a \cdot H$ for some $a>1$, we can randomly hash raw IDs to embedding table rows, $f_{\text{RH}}(x) := h(x)$, where $h(x): [1..I] \to [1..H]$ is one of the standard hash functions such as modulo hash. This creates random collisions with an average collision factor of $a$.

\paragraph{Semantic ID} The items' content embeddings are obtained from a multimodal image and text foundation model. The foundation model is pre-trained on a large training set of items using image and text alignment objectives \citep{radford2021learning}. The RQ-VAE is then trained on the content embeddings of all target items from the past three months, with $L=3$ and $K=2048$. We use the prefix $3$-gram paramaterization $f_{\text{SemId}}=p \circ s$ from \Cref{sec:token_parameterization}.

\vspace{.5em}
The analyses in \Cref{sec:segment_analysis} and \Cref{sec:representation_space} focus on the target item sparse feature. We train three versions of the recommendation model using the above three embedding lookup functions. The size of the item embedding table is equal to the total number of items for IE and set to a smaller size for RH and SemID, with the average collision factor of $3$. The user history features are mapped using random hashing. 

In \Cref{sec:user_history_modeling}, we use Semantic ID for the user history features and study its effect on aggregation module architectures. The item interaction history sequence length is fixed to $O(100)$. We pad or truncate the user history to fit the desired length.
%with padding items (zero embeddings) added if the number of interaction items is less than the maximum sequence length.
%The target item feature is also remapped with Semantic ID.

%For our experiments, the sequence of codes are mapped to the embedding table via a 3-layer prefix-gram parameterization. %, e.g., a code $[a_1, a_2, a_3, a_4]$ becomes $[a_1, (a_1+1)\cdot c+a_2, (a_1+1)\cdot c^2+(a_2+1)\cdot c+a_3]$. 
% Ablation studies on alternative parameterizations showed that prefix-gram slightly outperformed a bigram parameterization. After the mapping to the embedding table, the multiple embeddings are sum-pooled into a single embedding.

\subsection{Segment Analysis}
\label{sec:segment_analysis}
To understand the effect of impression skew on each approach, we segment the data based on the item's number of impressions in the training period. We  sort all items by impression count. As before, we segment items into head, torso, and tail items according to whether they generate $25\%$, $75\%$, or $100\%$ of the cumulative impression count in this sorted order. Due to the impression skew, the percentage of items that are head, torso, or tail items are $0.1\%$, $5.5\%$, or $94.4\%$, respectively. We also evaluate on the segment of new items that only appear in the evaluation period and were not seen during training. The performance of the three item representation approaches is shown in \Cref{tab:segment_analysis_generalization}.

% We expect Semantic ID to improve generalization for tail items, because individually, they do not have enough training examples for the model to learn good embedding representations. 
%To measure generalization, we evaluate the trained models on the examples from the next six hours after the training period. 

% We evaluate the trained models on three time periods: the next 6 hours after the training period, to measure generalization; the last 6 hours of training, to measure overfitting; and a middle 6-hour block of training (last 48 to 42 hours), to understand how ID drifting affects model performance. Only the last time period consists of held-out data. The results for Semantic ID, random hashing, and individual embeddings on different item segments and time periods are in \Cref{tab:segment_analysis}.

Compared to the baselines, Semantic ID improves generalization for tail items, is NE neutral for head items, and slightly beneficial for torso items. As this is also relative to the individual embeddings approach, Semantic ID is not simply better at clustering than random hashing, but we find that the target item feature benefits from semantics-based knowledge sharing.

Specifically, the knowledge sharing occurs through the shared embedding weights, which receive updates for semantically-related items. Knowledge sharing benefits are largest on the new items segment, where SemID achieves large gains over both RH and IE ($-0.41\%$ and $-0.33\%$, respectively). New items use pre-trained weights from semantically similar items seen during training for prediction, rather than using a non-relevant weight (RH) or an untrained weight (IE). 

% We additionally measure overfitting for each of the approaches. We compute a metric we call the \textit{generalization gap}, defined to be performance on the evaluation set (the next 6 hours after the training period) minus the post-trained model's performance evaluated on the last 6 hours of training data. A smaller generalization gap indicates less overfitting. The results are in \Cref{tab:segment_analysis_train_eval_gap}.

% \begin{table}[h]
%   \caption{Generalization gap: NE(Eval) - NE(Train[6h-0h]). Lower is better.}
%   \label{tab:segment_analysis_train_eval_gap}
%   \centering
%   \begin{tabular}{c|ccc}
%     %& \multicolumn{3}{c}{NE(Eval) - NE(Train)} \\ 
%     Cum. Exs. & RH & IE & SemID \\ \midrule
%     25\% (Head) & 0.0017 & 0.0017 & 0.0016 \\
%     75\% (Torso) & 0.0123 & 0.0121 & 0.0114 \\
%     100\% (Tail) & 0.0201 & 0.0203 & 0.0184 \\ \midrule
%     All Items & 0.0125 & 0.0124 & $\bm{0.0115}$ \\ \bottomrule
%   \end{tabular}
% \end{table}

% The generalization gap is larger for segments with less training examples, as expected. Semantic ID has the smallest generalization gap on all segments and benefits the tail segment most. % Interestingly, on tail items, individual embeddings overfits the most out of the three approaches. Semantic ID has the smallest generalization gap on all segments.

To measure the effect of embedding representation drifting on model performance, we evaluate the trained models back on the training data but on two different temporal segments. We take NE on 42-48 hours prior to the end of the training epoch and subtract NE on the last six hours of training. A smaller value indicates that the embedding representation shift caused by ID drifting affects model fit less. This is because our model is trained time-sequentially for one epoch, so the resulting model learns to fit the latest training time period at the end of training. The results are in \Cref{tab:segment_analysis_drifting_gap}.
% \begin{table}[h]
%   \caption{Sensitivity to distribution shift: NE(Train[48h-42h]) - NE(Train[6h-0h]). Lower is better.}
%   \label{tab:segment_analysis_drifting_gap}
%   \centering
%   \begin{tabular}{c|ccc}
%     %& \multicolumn{3}{c}{NE(Eval) - NE(Train)} \\ 
%     Cum. Exs. & RH & IE & SemID \\ \midrule
%     25\% (Head) & 0.0057 & 0.0065 & 0.0059 \\
%     75\% (Torso) & 0.0087 & 0.0075 & 0.0076 \\
%     100\% (Tail) & 0.0128 & 0.0103 & 0.0106 \\ \midrule
%     All Items & 0.0083 & 0.0074 & $\bm{0.0073}$ \\ \bottomrule
%   \end{tabular}
% \end{table}

Individual embeddings approach has a smaller performance gap compared with random hashing. This highlights that random hashing suffers from ID drifting -- the embedding representations lose capability to represent older items over time as the weights are updated using new item examples. In contrast, Semantic ID matches the performance of individual embeddings, indicating that its learned representations are more stable over time.
% SemID is able to generalize to earlier in the training period.
%This may be because some learning from the old items is preserved in the weights in addition to generalizing from more recent items, although we did not measure the two effects separately.

We conjecture that this improved representation stability also allows the model to generalize better over much longer training durations, where ID drifting becomes even more pronounced. 
% Does Semantic ID retain some knowledge from older training items in its learned representations, or is it simply generalizing better than random hashing? To answer this question, 
We train the RH and SemID models over a 20-day period and compare them to corresponding models trained only for the last four days of the period. 
% The longer the training period, the more severe ID drifting issue we have during training. 
The results in \Cref{tab:long_retention_exp} show that compared to random hashing, the performance of Semantic ID  scales better with training data over a longer period.
% , indicating that its representations retain learning from older items. 
\begin{table}[h]
  \caption{NE improvement from training for 20 days of data instead of 4 days.}
  \label{tab:long_retention_exp}
  \centering
  \begin{tabular}{ccc}
    \toprule
     &   RH  & Semantic ID \\ \midrule
     Eval NE Gain & $-0.18\%$  & $-0.23\%$ \\  \bottomrule
  \end{tabular}
\end{table}

\subsection{Item Representation Space}
\label{sec:representation_space}
To gain a better understanding of the item embedding representations, we extract the learned embedding weights from each trained model. One can view random hashing and Semantic ID as two different ways to partition the raw item ID corpus. We wish to see whether the semantics-based partition produced by Semantic ID is better suited for the recommendation problem  than the random partition produced by random hashing. 

When several items are assigned to the same partition, they get mapped to the same embedding by the embedding lookup module. We view this embedding vector as a summary of the per-item embeddings learned by the individual embeddings model. While we fit the individual embeddings model for illustration purposes in this paper, IE is impractical in industry-level settings. A partition with lower intra-partition embedding variance and higher inter-partition distances can be viewed as a more effective summary of individual embeddings. We compute these metrics for the RH and SemID partitions of the embeddings learned by the IE model.

We set the collision factor to $5$ for this experiment. As a result, the resulting clusters for RH and SemID partitions contain $5$ items on average. However, since Semantic ID is the latent codes learned by an RQ-VAE model, the resulting cluster sizes are highly variable. 
%For Semantic ID, we use the item's first three Semantic ID codes as the cluster ($[c_1,c_2,c_3]$). 
We compute metrics for two groups of Semantic ID clusters, small clusters with 4-10 items each and the top 1,000 clusters, where each cluster contains thousands of items. \Cref{tab:cluster_analysis} contains the average variances and average pairwise distances, with standard deviations in parentheses. The metrics are averaged across the embedding dimensions to produce single scalars for comparison.

\begin{table}[h]
  \caption{Intra- and inter-cluster variances and pairwise distances for random hashing and SemID-based partitions.}
  \label{tab:cluster_analysis}
  \centering
  \footnotesize
  \begin{tabular}{ccc}
    \toprule
     & Variance & Pairwise distance \\ \midrule
    \scriptsize{Random Hashing}  & \num{1.52e-3} \scriptsize{(\num{8.0e-4})}  & $0.22$ \scriptsize{($0.04$)}  \\
    \scriptsize{SemID (small)}   & \num{1.31e-3} \scriptsize{(\num{1.0e-3})}   & $0.24$ \scriptsize{($0.09$)}   \\
    \scriptsize{SemID (top 1,000)}   & \num{1.23e-3} \scriptsize{(\num{5.5e-4})}   & $0.06$ \scriptsize{($0.02$)}   \\ \bottomrule
  \end{tabular}
\end{table}

We observe that the Semantic ID partitions produce clusters with lower intra-cluster variance compared to random hashing. However, the resulting pairwise distances send a mixed signal. We hypothesize that the low pairwise distances between the top 1,000 clusters is because RQ-VAE places multiple centroids into the regions with highest data density to minimize the model loss.

\subsection{User History Modeling}
\label{sec:user_history_modeling}

In this section, we explore the effect of Semantic ID on user history modeling. One role of the module is to contextualize and summarize the user history. 
% For example, if a user's item interaction history contains a coconut water product, we could use the rest of the user's item history to distinguish whether more generally, the user likes flavored water or coconut-based products.

We find that using Semantic ID and a contextualizing attention-based aggregation module (PMA or Transformer) brings outsized gains compared to a baseline that does not contextualize the sequence (Bypass). These results are summarized in \Cref{tab:user_history_ne}. %where train NE is the window NE computed during training over the last 5B examples, and evaluation NE is computed from the examples in the next six hours following training.

\begin{table}[h]
  \caption{Performance for three aggregation modules. Baseline: model with RH for each module. Semantic ID brings larger gains to the contextualizing modules.}
  %Each baseline represents user history features with RH, and is compared to a model that uses Semantic ID.
  \label{tab:user_history_ne}
  \centering
  \begin{tabular}{cccc}
    \toprule
     & Train NE Gain & Eval NE Gain \\ \midrule
    Bypass  &  $-0.056\%$ & $-0.085\%$  \\
    Transformer & $-0.071\%$ & $-0.110\%$  \\
    PMA & $-0.073\%$ & $-0.100\%$  \\ \bottomrule
  \end{tabular}
\end{table}
% \begin{table}[h]
%   \caption{Train NE for different aggregation modules when representing user history features with either random hashing or Semantic ID. Semantic ID shows improvements for all three aggregation modules, but brings larger gains to the contextualizing modules.}
%   \label{tab:user_history_ne}
%   \centering
%   \begin{tabular}{cccc}
%     \toprule
%      & RH & SemID & vs. RH Gain \\ \midrule
%     Bypass  & $0.8269$ & $0.8264$ & $0.056\%$  \\
%     Transformer & $0.8268$ & $0.8262$ & $0.071\%$ \\
%     PMA & $0.8268$ & $0.8262$ & $0.073\%$  \\ \bottomrule
%   \end{tabular}
% \end{table}

To understand how using Semantic ID changes the learned attention patterns in PMA and Transformer aggregation modules, we compute four metrics on the attention scores on a random subset of 1,000 evaluation examples. 

Let $\mathbf{A} \in \mathbb{R}^{T\times S}$ be the attention score matrix, where $T$ is the target sequence length and $S$ is the source sequence length. We have $T=S$ for Transformer and Bypass, and $T=32$ for PMA. Each row $\mathbf{a}_{i,:}$ of $\mathbf{A}$ represents a probability distribution over the source tokens.
%($\mathbf{a}_{:,i}$ lies on the $(S-1)$-simplex).
%Let $\mathbf{a}_i$ be the $i$-th row of $\mathbf{A}$. We have $\mathbf{a}_i \in \Delta^{S-1}$, where $\Delta^{S-1}$ denotes the $(S-1)$-simplex.
The metrics we consider are defined as follows.
\begingroup
\small
\begin{align*}
\addtolength\jot{-4pt} %controls line spacing
    &\textit{First source token attention:} &  &{\textstyle  \frac{1}{T} \sum_{i=1}^T a_{i,1}} \\
    &\textit{Padding token attention:} & &{\textstyle \frac{1}{T} \sum_{i=1}^T \sum_{j=1}^S \mathbb{I}{\{a_{i,j} =\text{pad}\}} \cdot a_{i,j}} \\
    &\textit{Entropy:} & &{\textstyle \frac{1}{T} \sum_{i=1}^T \sum_{j=1}^S a_{i,j} \cdot \log_2{a_{i,j}}}\\ 
    &\textit{Token self-attention:} & &\textstyle{\frac{1}{T} \sum_{i=1}^T a_{i,i}}
\end{align*}
\endgroup

% \noindent \textit{First source token attention:} $\frac{1}{T} \sum_{i=1}^T s_{i,1}$

% \noindent \textit{Padding token attention:} $\frac{1}{T} \sum_{i=1}^T \sum_{j=1}^S \mathbb{I}{\{s_{i,j} \text{ is a pad token}\}} \cdot s_{i,j}$

% \noindent \textit{Entropy:} $\frac{1}{T} \sum_{i=1}^T \sum_{j=1}^S s_{i,j} \cdot \log_2{s_{i,j}}$

% \noindent \textit{Token self-attention:} $\frac{1}{T} \sum_{i=1}^T s_{i,i}$

%\vspace*{-0.2cm}
\begin{table}[h]
  \caption{Attention score-based evaluation metrics for random hashing and SemID-based models for the user history item interaction features.}
  \label{tab:user_history_attention_metrics}
  \centering
  \begin{tabular}{ccccc}
    \toprule
     & First & Pad & Entropy & Self \\ \midrule
    Transformer + RH & $0.030$ & $0.460$ & $2.149$ & $0.052$ \\
    Transformer + SemID & $0.043$ & $0.418$ & $1.967$ & $0.045$  \\ \midrule 
    PMA + RH  & $0.071$ & $0.351$ & $3.075$ & --  \\
    PMA + SemID & $0.074$ & $0.313$ & $3.025$ & -- \\
    \bottomrule
  \end{tabular}
\end{table}
% We also visualize the attention scores sorted by magnitude, averaged over the evaluation examples.
% \cz{todo viz}
From the metric readings in \Cref{tab:user_history_attention_metrics}, we see that models trained with Semantic ID have lower entropy, token self-attention, and padding token attention, and higher attention score on the first source token in the sequence. This means that Semantic ID-based models place more weight on higher-signal tokens (i.e., the first and most recent item in the sequence, rather than earlier and potentially stale tokens or padding tokens), have attention score distributions that are less diffuse over the entire sequence (i.e., lower entropy), and for Transformer, place higher weights on other tokens rather than self-attending. These metrics are promising signals that Semantic ID item representations are more stable and meaningful than their random hashing-based counterparts in user history modeling.

\section{Productionization}
The Semantic ID features have been productionized in Meta Ads Recommendation System for more than a year. They serve as top sparse features in the existing ads ranking models according to feature importance studies. In this section, we provide an overview of the online serving pipeline and key implementation details.

\subsection{Offline RQ-VAE Training}
The RQ-VAE models are trained on Content Understanding (CU) models for ads ranking at Meta. The CU models are pre-trained on the public CC100 dataset \cite{conneau2019unsupervised} and then fine-tuned on internal ads datasets. We sample ad IDs and their corresponding content embeddings from the past three months' data and train the RQ-VAE model offline. For production models, we train RQ-VAEs with $L = 6$ and $K= 2048$, and Semantic ID follows the design of prefix-5gram from \Cref{sec:token_parameterization} with $H=O(50M)$. After training, we use a frozen RQ-VAE checkpoint for online serving.
%The temporal stability of RQ-VAE model was studied in \cite{singh2023better}. We observe similar properties in Meta Ads ranking.

\subsection{Online Semantic ID Serving System}
\begin{figure}[t]
    \centering
    \includegraphics[width=\linewidth]{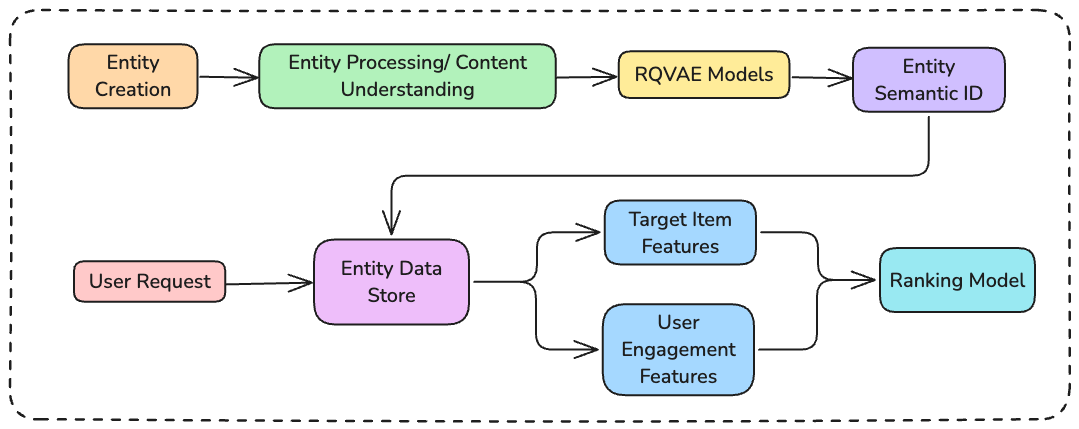}
    \caption{Semantic ID serving pipeline.}
    \label{fig:online}
\end{figure}
\Cref{fig:online} shows the online serving pipeline of real-time Semantic ID features. At ad creation time, we process the ad content information and provide it to the CU models. The output CU embeddings are then passed through the RQ-VAE model which computes the Semantic ID signal for each raw ID. The signal is then stored in the Entity Data Store. At the feature generation stage, the  target item raw ID and user engagement raw ID histories are enriched with the Semantic ID signal from the Entity Data Store to produce semantic features. When serving requests arrive, the precomputed features are fetched and passed to the downstream ranking models.
%The whole feature generation process takes less than 10 seconds.

\subsection{Production Performance Improvement}
We created six sparse and one sequential feature from different content embedding sources, including text, image, and video, and report the NE gain on the flagship Meta ads ranking model in \Cref{tab:production_ne_gain}. In Meta ads ranking, an offline NE gain larger than $0.02\%$ is considered significant. Overall, across multiple ads ranking models, incorporating the Semantic ID features have yielded a $\mathbf{0.15\%}$ {\bf gain in performance on our top-line online metric}. As the Meta Ads recommender serves billions of users and has been one of the most optimized models in the company, a $0.15\%$ online performance gain is considered significant.

% Overall, the Semantic ID features have resulted in $0.15\%$ online performance gain across different models.
\begin{table}[h]
  \caption{NE improvement from incorporating Semantic ID features in the flagship Meta ads ranking model.}
  \label{tab:production_ne_gain}
  \centering
  \footnotesize
  \begin{tabular}{lccc}
    \toprule
     & \scriptsize{Train NE Gain} & \scriptsize{Eval NE Gain} \\ \midrule
     % Baseline  &  $0$ & $0$   \\ 
    Baseline + 6 sparse features  &  $-0.063\%$ & $-0.071\%$   \\ \midrule
    Baseline + 1 sequential feature  &  $-0.110\%$ & $-0.123\%$   \\ \bottomrule
  \end{tabular}
\end{table}

\subsection{Semantic and Prediction Similarity}
\label{sec:online_semantic_similarity}
Intuitively, one may think that if two items are semantically similar, their user engagement patterns will also be similar. However, user behavior and perceptions are more nuanced and not predictably continuous with respect to semantics. For robust delivery performance with Semantic ID, we must ensure a degree of continuity (or correlation) of the ranking model's behaviour with respect to the semantic similarity relation between the items in our system. 

To measure this correlation, we conduct an online A/B test where we select a set $S$ of items that are recommended to a user by our system. For a given user, with $50\%$ probability, we mutate the set $S$ to $S'$ by randomly swapping an item in $S$ with a different item with the same prefix from Semantic ID. This operation results in a

\begin{equation}
 \textit{Click Loss Rate} := \frac{ \text{CTR on } S' -  \text{CTR on } S }{\text{CTR on } S}.
\end{equation}

\begin{figure}[h]
    \centering
    \includegraphics[width=\linewidth]{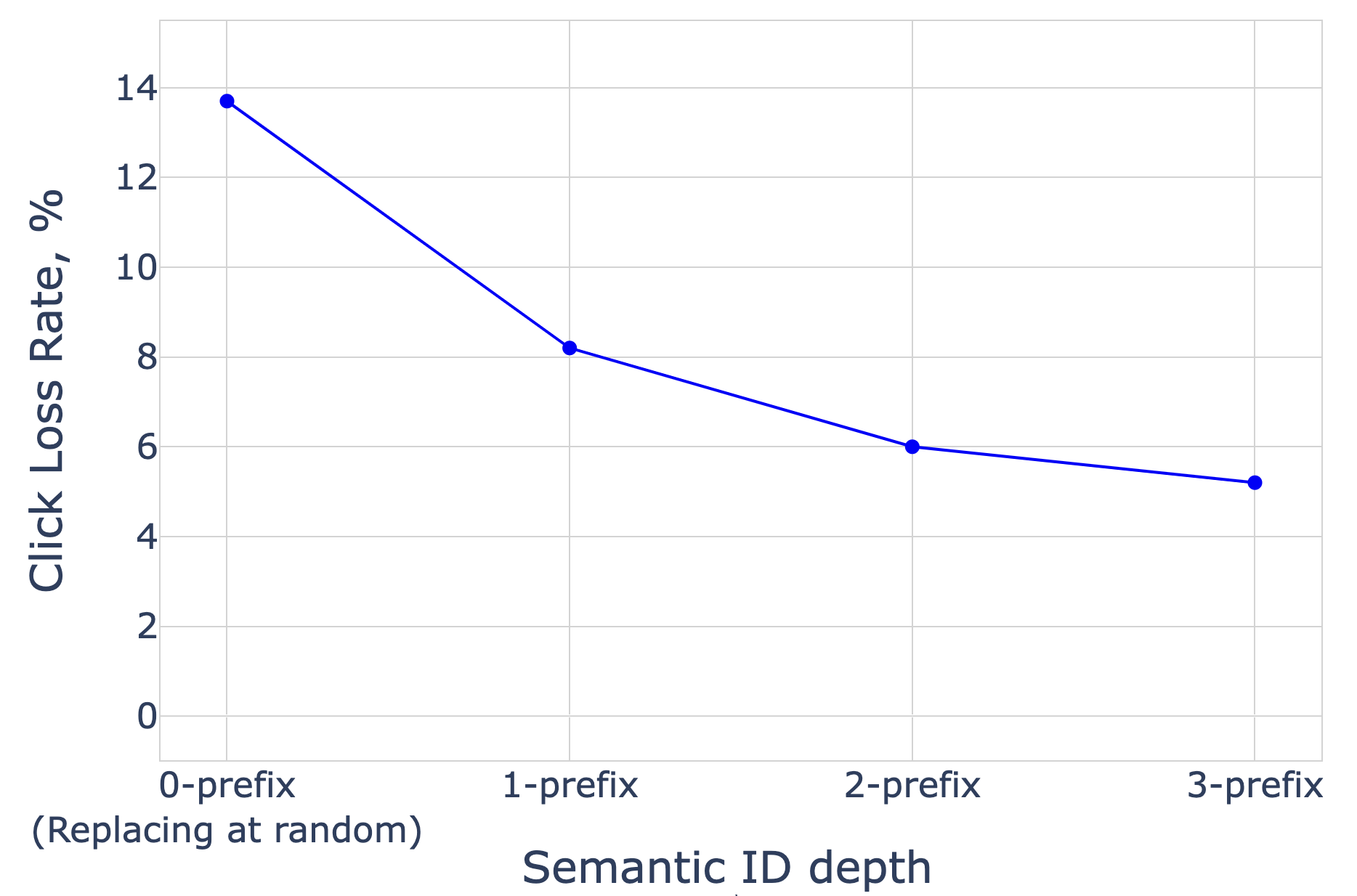}
    \caption{Click Loss Rate reduction from Semantic ID.}
    \label{fig:click_loss_reduction}
\end{figure}

The click loss rate decrease from using deeper prefixes from Semantic ID is summarized in \Cref{fig:click_loss_reduction}. 

% \begin{table}[h]
%   \caption{Click Loss Rate Reduction from Semantic ID}
%   \label{tab:click_loss_reduction}
%   \centering
%   \begin{tabular}{cccc}
%     \toprule
%      & Click Loss Rate \\ \midrule
%     0-prefix Semantic ID (replacing at random)  &  $13.7\%$    \\ 
%     1-prefix Semantic ID  &  $8.2\%$   \\ 
%     2-prefix Semantic ID  &  $6.0\%$   \\
%     3-prefix Semantic ID  &  $\bm{5.2\%}$  \\ \bottomrule
%   \end{tabular}
% \end{table}

Since Semantic ID partitions the item corpus based on item semantics, we conclude that prediction similarity is correlated to semantic similarity. This supports the representation space analysis result in \Cref{sec:representation_space}. Moreover, the hierarchy of codes in Semantic ID effectively captures the finer-grained details of an item's semantics: deeper prefixes monotonically reduce the click loss rate.

%Note that a $3$-prefix does not completely resolve the corresponding semantic cluster to contain only the A or A' copies of an original item. The cluster will also contain more semantically distinct relatives of the original item. Nevertheless, the 3-prefix resolves the semantics enough to drastically reduce the Click Loss Rate.

\subsection{A/A Variance}

Yet another disadvantage of a ranking model based on random hashing is inherent model prediction variance, resulting in downstream ad delivery variance. Specifically, one can create a copy of an item with a different raw item ID. Then, both the original and item copy enter the recommendation system. Because the embeddings will be different for the original and item copy after hashing, the model predictions and delivery system behavior will also differ. We call this  phenomenon the {\it A/A variance}, where ``A/A'' signifies that we consider an exact copy of the original item.\footnote{A related type of a ranking model variance is the {\it A/A' variance} where one instead considers minor item differences instead of exact copies, such as two images with the same object in the foreground, but with backgrounds of different colors.} 
This variance is undesirable since it reduces the robustness of the downstream ad ranking orders and the system's ability to accurately target correct audiences. Semantic ID helps reduce the A/A variance by eliminating the randomness caused by random hashing -- exact copies or very similar items will often have the same $k$-prefix Semantic ID.

We set up an online shadow ads experiment where we measure the relative A/A prediction difference (AAR) for a given model. For an A/A pair $(a_1, a_2)$,
\begin{equation}
 \text{AAR}(a_1, a_2) := 2\frac{p(a_1) - p(a_2)}{p(a_1) + p(a_2) + \epsilon},
\end{equation}

\noindent where $p(a_i)$ is the ranking model prediction for item $a_i$. 

The production model with six Semantic ID sparse features achieves a  $\mathbf{43\%}$ {\bf reduction in average AAR} compared to the same model without the six features. We believe that the majority of AAR reduction is from the tail items, as studied in \Cref{sec:segment_analysis}.

\section{Conclusion}
We show how Semantic ID can be used to create a stable ID space for item representation and propose Semantic ID prefix-ngram, which significantly improves Semantic ID’s performance in ranking models. In offline experiments, we study trained ranking models and find that under Semantic ID, the harmful effects of embedding representation instability are mitigated compared to random hashing and individual embeddings baselines. We detail the successful productionization of Semantic ID features in Meta's ads recommendation system, and show that the online production system obtains significant performance gains as well as reduced downstream ad delivery variance.

% In this paper, we study the sources of gains from Semantic ID over the commonly used method of random hashing (RH) in large-scale recommender systems. We find that: Semantic ID clusters are more meaningful than RH clusters, Semantic ID improves generalization for torso and tail items (even compared to a baseline where each item receives its own embedding), and Semantic ID is less sensitive to distribution shift from ID drifting. In user history modeling, Semantic ID is able to more effectively aggregate sequential features. These findings deepen our understanding of SemID's sources of gains as beyond simply benefiting performance.

%\begin{acks}
%    The authors would like to thank Tao Liu, Chaofei Yang, Yiqun Liu, Daisy Li, Manish Keshri, Aryan Pandhi, Qingyi Lu, Hangjun Xu, Xichan Liu, Chenye Zhao, Weilin Zhang, Birmingham Guan, Austin Zhao, Chris Cadonic, Esam Abdel Rhman, Shakti Kumar, Wen-Yen Chen, Yiping Han, Jie Zheng, Subbu Subramanian, John Bocharov, Bo Long, Wenlin Chen, Sri Reddy, Rocky Liu, Santanu Kolay, Sandeep Pandey and others who contributed, supported and collaborated with us.
%\end{acks}

\bibliographystyle{ACM-Reference-Format}

\appendix
\section{Aggregation Module Architectures}
\label{app:aggregation_modules}

% \section{Parameterization Ablations}
% \label{app:sem_id_parameterizations}

%\section{Calibration-Free Normalized Entropy}
%\label{app:calibration_free_ne}

\noindent \textit{Bypass.} Apply a linear weight matrix $\mathbf{W} \in \mathbb{R}^{d_m \times d_m}$ to each embedding separately,
\begin{equation}
    \text{Bypass}(\mathbf{X}) := \mathbf{X}\mathbf{W}.
\end{equation}

\noindent \textit{Transformer} \citep{vaswani2017attention}. Apply a Transformer layer to the embedding sequence. The attention submodule is defined as
\begin{align}
    \text{Attention}(\mathbf{X}) &:= \text{softmax}\left(\frac{(\mathbf{X}\mathbf{W}^Q)(\mathbf{X} \mathbf{W}^K)^\intercal}{\sqrt{d_m}}\right) (\mathbf{X}\mathbf{W}^V),
    \label{eq:trf_submodule}
\end{align}
where $\mathbf{W}^Q$, $\mathbf{W}^K$, $\mathbf{W}^V \in \mathbb{R}^{d_m \times d_a}$ are the query, key, and value weight matrices, respectively, and $d_a$ is the query/key/value vector dimension. The full Transformer module is given by
\begin{align}
    \mathbf{X}^{(1)} &= \text{Attention}(\text{LayerNorm}(\mathbf{X})) + \mathbf{X}\label{eq:trf_full_module_1} \\
    \mathbf{X}^{(2)} &= \text{MLP}(\text{LayerNorm}(\mathbf{X}^{(1)})) + \mathbf{X}^{(1)},
    \label{eq:trf_full_module_2}
\end{align}
where LayerNorm and MLP designate the standard layer norm and position-wise MLP layers. We add standard positional embeddings to the encoding before applying Transformer or PMA modules.

\noindent \textit{Pooled Multihead Attention (PMA)} \citep{lee2019set}. Apply a Transformer layer to the embedding sequence, but replace the attention query vectors with $d_s$ learnable weight vectors. The PMA attention submodule is defined as
\begin{equation}
    \text{PMAttention}(\mathbf{X}) := \text{softmax}\left(\frac{\mathbf{S} (\mathbf{X}\mathbf{W^K})^\intercal}{\sqrt{d_m}}\right)(\mathbf{X}\mathbf{W}^V),
    \label{eq:pma_submodule}
\end{equation}
where $\mathbf{S} \in \mathbb{R}^{d_s \times d_a}$ is comprised of $d_s$ learnable query vectors, or seeds. In our experiments, $d_s=32$.

The PMA module is formed using the same equations as for the Transformer module (Equations \ref{eq:trf_full_module_1} and \ref{eq:trf_full_module_2}), except that PMAttention is used in place of Attention.

\section{Distributions of Clicks for Semantic ID}
\label{app:semid_dist}
% Here we illustrate how the click distribution in the raw ID space changes when we pass to the Semantic ID space.
\begin{figure}[h]
    \centering
    \includegraphics[width=\linewidth]{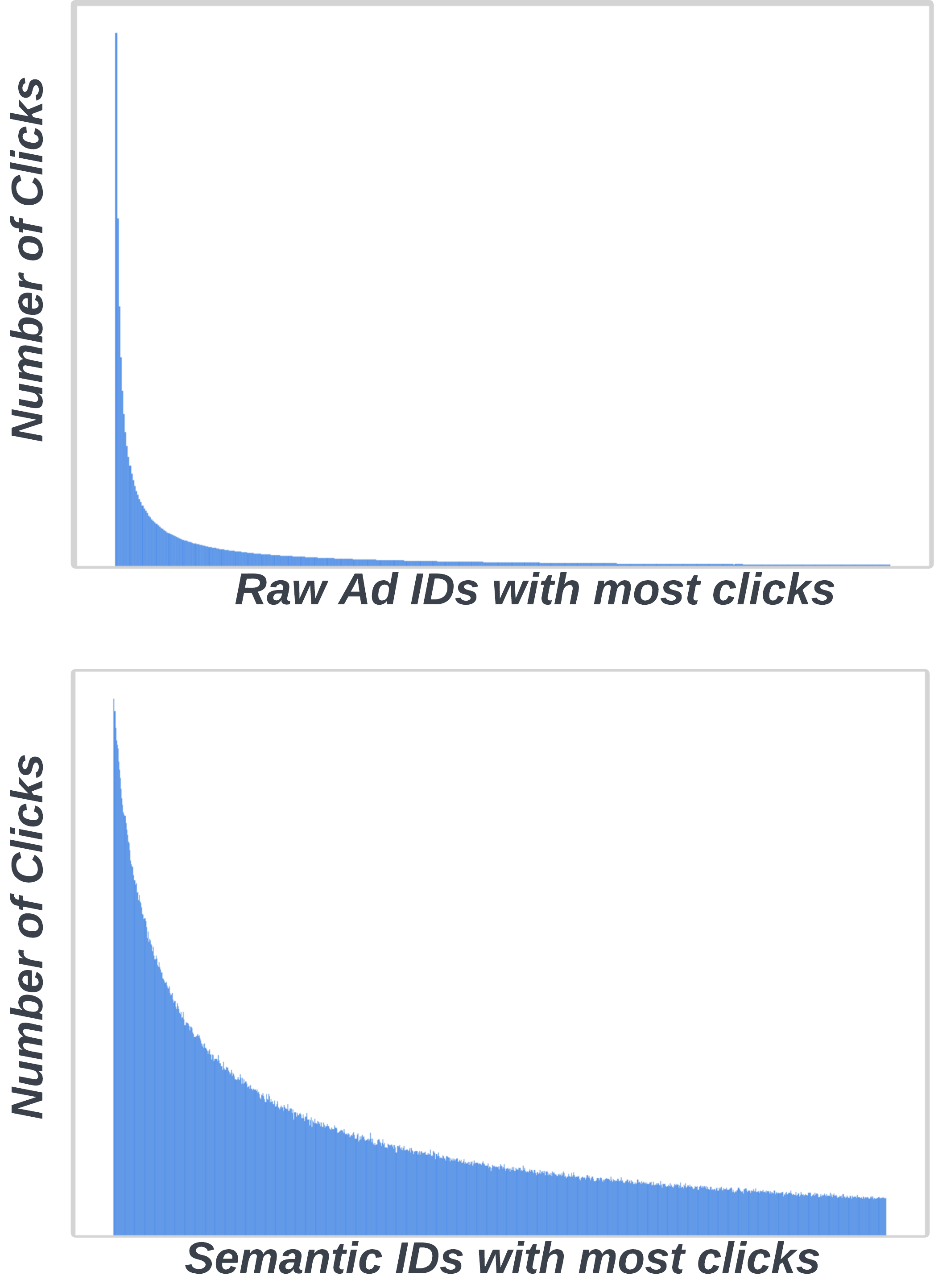}
    \caption{The 30-day click distribution in raw ID and Semantic ID spaces.}
    \label{fig:semid_dist}
\end{figure}

The click distribution in Semantic ID space (\Cref{fig:semid_dist}) clearly exhibits less skew compared to the click distribution in the raw ID space. Note that while \Cref{fig:impr_skew} shows the cumulative distribution of impressions, \Cref{fig:semid_dist} shows the the marginal distribution of clicks.

\end{document}